\documentclass[conference, a4paper, 10pt]{IEEEtran}
\usepackage[T1]{fontenc}
\usepackage[utf8]{inputenc}
\usepackage[cmyk]{xcolor}
\usepackage{multirow}

\usepackage{makeidx}
\usepackage{tabularx}
\usepackage{verbatim}
\usepackage[final]{graphicx}
\usepackage[final]{listings}
\usepackage{framed}

\usepackage{amsthm}
\usepackage{ifdraft}
\ifdraft{
\usepackage[firstpage]{draftwatermark}
\SetWatermarkScale{4}
}{}

\usepackage{todonotes}
\newtheorem{mydef}{Definition}

\begin{document}
%
\title{Reactive User Behavior and Mobility Models}

\author{\IEEEauthorblockN{Anna F\"orster, Anas Bin Muslim, Asanga Udugama}
\IEEEauthorblockA{University of Bremen, Germany\\
Sustainable Communication Networks Group\\
Email: [afoerster,adu]@comnets.uni-bremen.de}
}


%


\maketitle

\begin{abstract}
In this paper we present a set of simulation models to more realistically mimic the behaviour of users reading messages. We propose a User Behaviour Model, where a simulated user reacts to a message by a flexible set of possible reactions (e.g. ignore, read, like, save, etc.) and a mobility-based reaction (visit a place, run away from danger, etc.). We describe our models and their implementation in OMNeT++. We strongly believe that these models will significantly contribute to the state of the art of simulating realistically opportunistic networks.
\end{abstract}


\IEEEpeerreviewmaketitle


\section{Introduction}\label{sec:intro}

The research area of opportunistic and device-to-device communications has seen a very large growth in recent years. Applications and services are arising all over the world and begin to offer a real alternative to infrastructure-based ones. However, their testing and evaluation is extremely tedious and complex, as it relies on end users, on their behaviour and mobility patterns. Researchers have been using simulation for this purpose for a long time, but the existing simulation models are very restricted in mimicking real users. 

Current user mobility models assume that people move in their environment without taking into consideration the data they receive through the network. Let us explore one of the mostly used motivational scenarios for opportunistic networks: disaster alert. When an accident or a natural disaster happens, like a fire or an earthquake, the application under test is supposed to send alert messages via device-to-device communications to all people around to warn them. In a real network, the reception of such a message will result in people running away. In a state-of-the-art simulation people will continue moving around as nothing has happened. This is the first problem we target with this work:

\textbf{Goal 1: Users should react to the application messages in an appropriate way and change their moving pattern.}

Furthermore, in current simulations, the user is assumed to produce and receive some messages without any meaning. In some cases the user provides preferences and wants to receive only a subset of all messages. However, the behaviour is also highly deterministic, which does not correspond at all to real human behaviour. This is our second goal:

\textbf{Goal 2: Give meaning to the messages exchanged and provide the simulated user with an ability to react to these messages and to act non-deterministically.}



In this paper, we propose a suit of simulation models to target the above problems. They are implemented as part of the new Opportunistic Protocol Simulator (OPS) in OMNeT++\footnote{https://github.com/ComNets-Bremen/OPS}\cite{udugama:2017ops}.
Section~\ref{sec:user} describes our user behaviour model. Section~\ref{sec:appl} focuses the parameterisation of our model for various application scenarios. Section~\ref{sec:mob} describes the corresponding reactive mobility model, while Section~\ref{sec:metrics} discusses the metrics to be used with our model for evaluating opportunistic networks. Section~\ref{sec:impl} describes the implementation of the models under OMNeT++, while Section~\ref{sec:conc} finally concludes the paper.





\section{Reactive User Behavior Model}\label{sec:user}

Our model is based on the idea that the user reacts to all messages she sees in the system, e.g. she might ignore or delete the message, she might "like it", she might additionally comment on it or she might even save it for later reference. Which reaction the user has depends on the application, on the type of messages, on the number of messages and on the interests of the user. It also depends on the level of activity of this particular user - some users tend to comment on many messages, some react only sporadically.

\begin{mydef}
A \textsc{user} is defined as a tuple $u = (INT, R, base)$, where $INT = \{i_1, ... i_m\}$ are the $m$ interests of this user, stored as keywords, $R= \{r_1, ... r_n\}$ are the possible reactions of the user and $base = Pr[X = r_i]$ is the probability of this user to select a particular reaction $r_i$.
\end{mydef} 

With other words, the user is identified by her interests, her possible reactions to messages and her general attitude of reacting to messages. The $base$ probability should be a heavy tail distribution, e.g. a log-normal distribution. Such distributions have been shown many times to model well the behaviour of people, e.g. how many friends they have~\cite{newman:2010}. Examples are provided in Section~\ref{sec:appl}.
The possible reactions are assumed to be an ordered set, where the first reaction is typically ignoring/deleting a message, while the next ones represent "heavier" reactions. For example, the user could vote (like) a message or save it for later. The last reaction is assumed to be the maximally possible reaction.

\begin{mydef}
A \textsc{message} is defined as a tuple $msg = (KEYS, pop, start, end, addr, radius)$, where $KEYS = \{k_1 ... k_l\}$ are the keywords associated with this message, $pop \in [0 .. 100]$ is the popularity of this message in the complete network (also defined as percentage of people who will react maximally at it), $start$, $end$ and $addr$ are used for messages carrying information about events. The radius is used to represent the danger area for emergency messages.
\end{mydef} 

Note that all parameters of $msg$ are optional. In the extreme case $msg$ does not carry any meaning nor information, which does not make sense in the real world. Thus, either $KEYS$ or $pop$ or $start,end,addr$ should be present. Only emergency messages have the $radius$ field.

\begin{figure}[t]
\begin{center}
\includegraphics[width = \columnwidth]{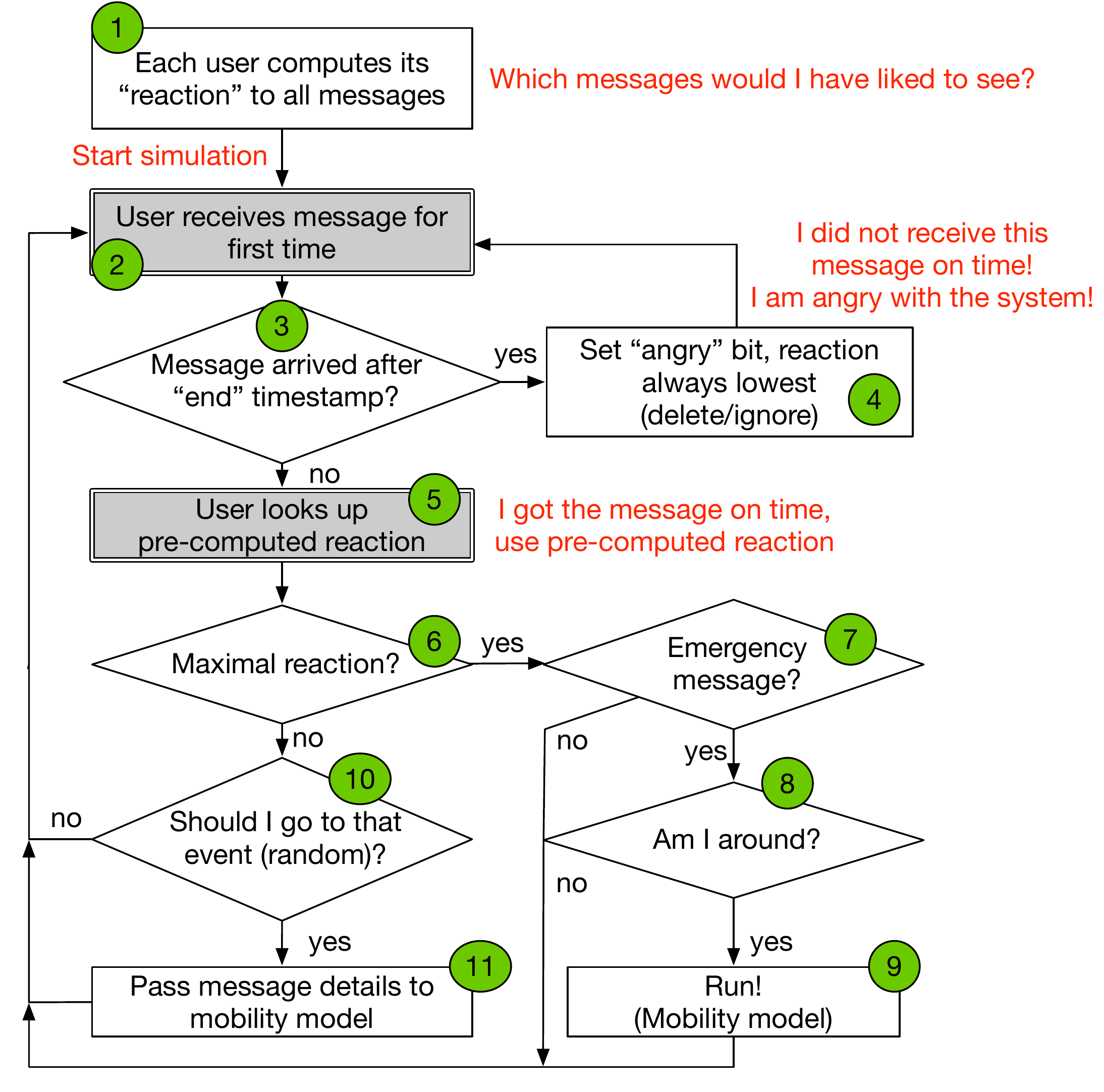}
\vspace{-6mm}
\caption{General approach to compute and use user reactions.}
\vspace{-6mm}
\label{fig:ground}
\end{center}
\end{figure}

The general flow of our model is depicted in Fig.~\ref{fig:ground}. Note that the data dissemination protocol itself is not part of it. What kind of information it is using and how remains out of scope of this paper. In the next sub-sections, we will focus on the individual steps from Fig.~\ref{fig:ground}.

\subsection{Pre-Computing Reactions to Messages}
The intuition behind our model is that there will be messages, which the user receives on time and takes some decision what to do and there will be messages, which arrive too late and are of no use any more. However, the user might have been interested in them (e.g., missed a concert of the favourite band). In real user experiments, we would ask the user \textit{after} the experiment: Which messages would you have liked to see? For simulation efficiency purposes, we pre-compute these interests before we start the simulation and to look them up later. 

The computation itself (step 1 in Fig.~\ref{fig:ground}) depends on three parameters: the base activity level of the user, the popularity of a particular event and the matching keywords between the user and the individual event. While the exact weight of these parameters can be changed and fitted to real-world applications and users' behaviour, the general assumption is that a user would react more significantly to more popular messages (e.g. good jokes) and to events which better match her interests.

\begin{figure}[htbp]
\begin{center}
\includegraphics[width = 0.7\columnwidth]{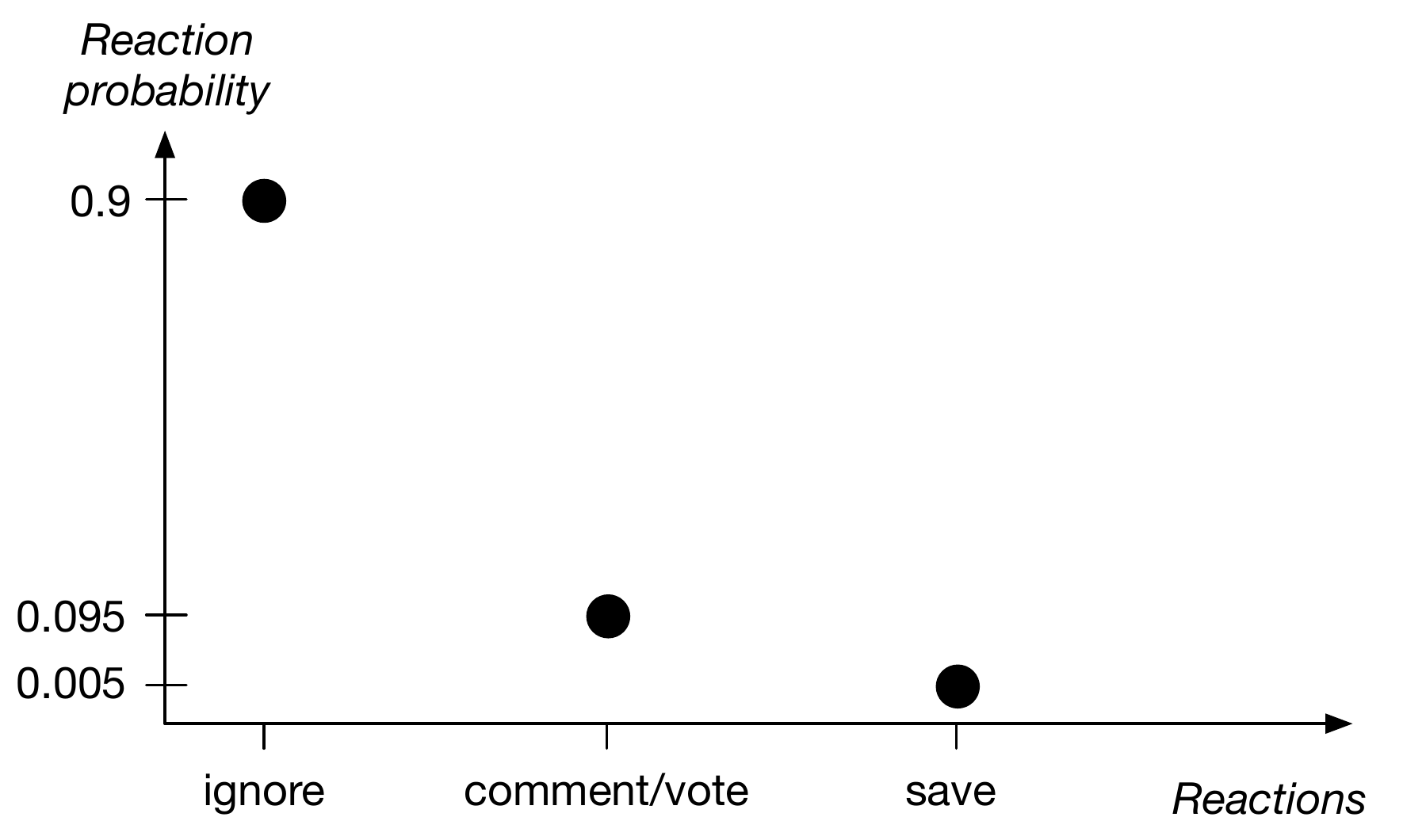}
\vspace{-2mm}
\caption{Sample base activity probability distribution of a user. There are three possible reactions in total and most of the messages get ignored.}
\vspace{-6mm}
\label{fig:base}
\end{center}
\end{figure}

In our model, each reaction is associated with a selection probability: the base activity of the user. An example is provided in Fig.~\ref{fig:base}. There are three possible reactions, \textsc{ignore, vote/comment, save} and the user ignores 90\% of her messages, comments/votes on 9.5\% of them and saves only 0.5\%. This example is based on observations of real user data exchange of the Jodel application~\footnote{https://jodel-app.com} over 3 days in Bremen, Germany.

\begin{figure}[htbp]
\begin{center}
\includegraphics[width = 0.7\columnwidth]{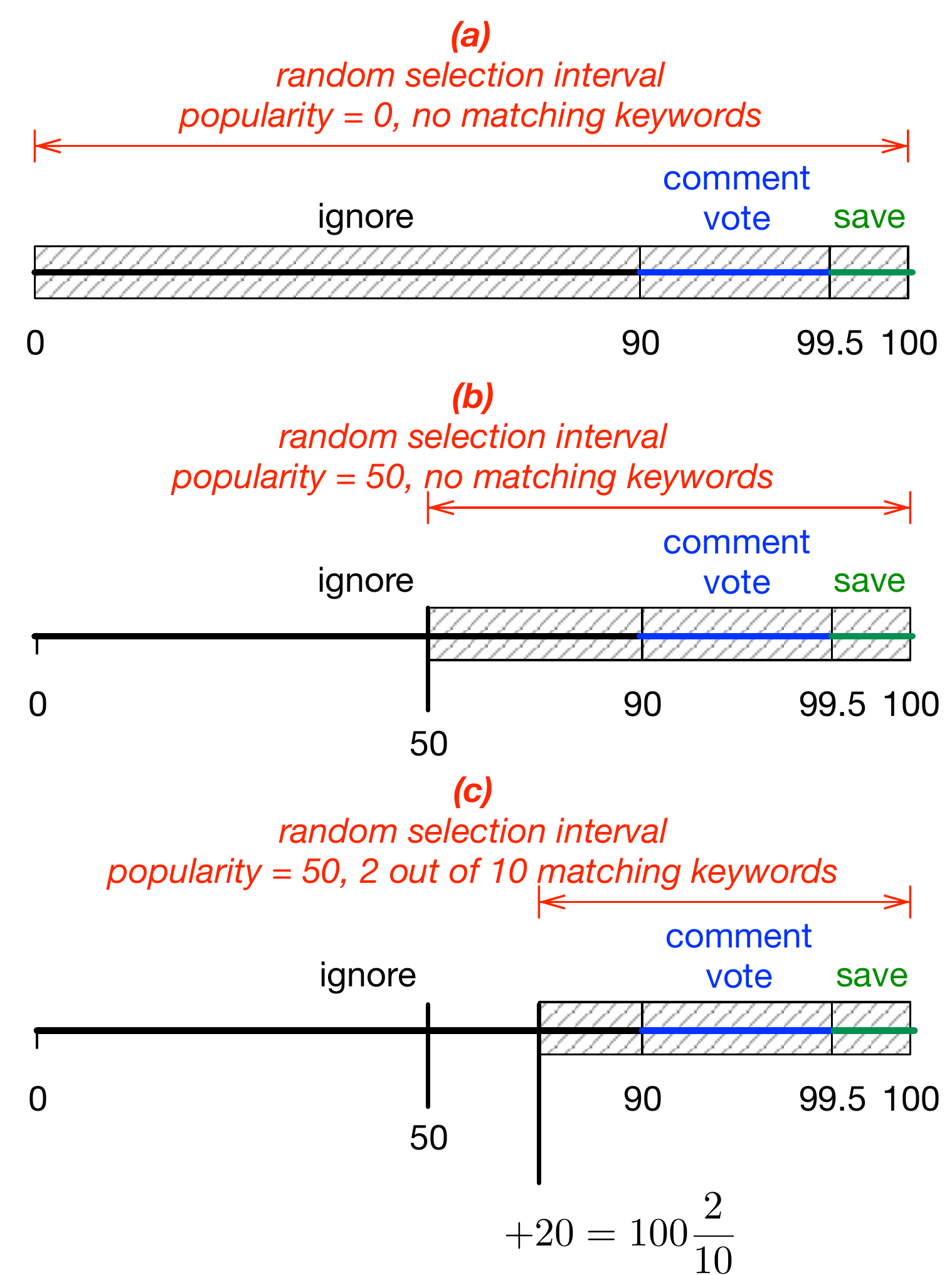}
\vspace{-2mm}
\caption{Selecting the user's reaction for a message.}
\vspace{-5mm}
\label{fig:intervals}
\end{center}
\end{figure}

If the popularity of the message is zero and there are no matching keywords with the interests of this user, the base probability is used directly to compute the reaction: in our example, the user will decide with 90\% probability to ignore the message. If we represent these probabilities as differently spaced intervals and use a simple uniform random number generator, the intervals will look as in Fig.~\ref{fig:intervals}(a) and the random number is computed as:

\begin{equation}\label{eq:rand}
r_{msg}^{user} = rand(0,100)
\end{equation}

In Fig.~\ref{fig:intervals}(a), the interval from which we can draw our random number is shaded and the probabilities of the individual reactions correspond exactly to the base probability of the user.

If the message is more popular, we need to give preference to "stronger" reactions. We do this by limiting the interval, from which we can draw our random number, as shown in Fig.~\ref{fig:intervals}(b). This can be computed as:

\begin{equation}\label{eq:rand_pop}
r_{msg}^{user} = rand({pop_{msg}}, 100)
\end{equation}

where $pop_{msg}$ is the popularity of this event. In this way, we limit the probability of selecting the lowest reaction and the probabilities of selecting one of the stronger reactions grow.

Furthermore, if there are matching keywords between the event and the interests of the user,  we limit the possible selection interval further:
\begin{equation}\label{eq:rand_full}
r_{msg}^{user} = rand({pop_{msg}} + \frac{100 k_{msg}^{user}}{l_{msg}}, 100)
\end{equation}

where $k_i^{user}$ is the number of matching keywords between this message and this user and $l_{msg}$ is the total number of keywords of this message. 


\subsection{Mobility Reactions}

The user could also react to a message with movement. For example, she could decide to go a concert or she might need to run away from a danger area.
We consider two possible mobility reactions:

\textbf{Immediate move:} when the user receives a disaster alert and is in the area of danger (provided by the $radius$ field), she immediately moves out of the danger area. 
The exact location to which she moves is handled by the mobility model.

\textbf{Scheduled move:} upon reception of a message about a city event or similar, the user might decide to go there at the time of the event and passes this information to the mobility model.

As already noted in Section~\ref{sec:intro}, this mobility reactiveness of the user requires the mobility model to change immediately the movement pattern or to schedule a move to some place later on. We will discuss the implementation of this in Section~\ref{sec:mob}.


\subsection{Simulation}

Once the reactions of all users to all messages have been precomputed, we are ready to start the simulation of our User Behaviour Model. It follows closely the flow chart of Fig.~\ref{fig:ground}. For every message the user receives for the first time (step 2), the User Behaviour Model needs to react. It first checks whether this message has an $end$ parameter and whether the message was received before this timestamp (step 3). If not, we mark this message with a special "angry" bit to be used later for computing the delivery rate of the system (step 4). Additionally, we always react to those late messages with the lowest reaction available (ignore/delete). 

If the message arrived on time, we look up the pre-computed reaction (step 5). If that reaction was maximal (step 6), then we randomly decide whether the user will visit this event or not (step 10). If yes, $start, end, addr$ are passed to the mobility model to plan the move (step 11).
A special case is provided when emergency messages arrive - if the user is in the danger zone of the emergency, she needs to exit the danger area immediately. Thus, the mobility model is told to "run away" (steps 8-9).

In case $start, end, addr$ are missing from the message, the flow becomes very short and only steps 1-2-5 are followed.


\section{Application Model and Examples}\label{sec:appl}

In this section, we give some examples of applications and how to parametrise our user behaviour model accordingly. The proposed parameters are summarised in Table~\ref{tab:params}. While this table suggests rather simple parameter sets, e.g. the same reaction probability for all users, it is of course possible to represent also more complex scenarios, e.g. different reaction probabilities for some more active users. Whether such complex scenarios deliver more realistic results is unclear and we will explore it in our future research.

\begin{table*}[t]
\caption{Summary of some applications with their parameters for the user behavior model.}
\vspace{-6mm}
\begin{center}
\begin{scriptsize}
\begin{tabular}{| p{1.4cm} | p{0.7cm} | p{2.1cm} | p{2.4cm} | p{0.8cm} | p{1.1cm} | p{2.1cm} | p{1.4cm} | p{2.3cm} | }
\hline
\multirow{2}{*}{\textbf{Application}} & \multicolumn{3}{ c | }{\textbf{User}} & \multicolumn{5}{ c |}{\textbf{Messages}}\\
\cline{2-9}
& \textbf{Number} & \textbf{Interests} & \textbf{Reactions \& \newline Probability} & \textbf{Number} & \textbf{Traffic (creation)} & \textbf{Keywords} & \textbf{Popularity} & \textbf{Time and place} \\
\hline
Jodel-like \newline campus & 500-1000 & none &  Ignore (90\%), \newline comment/vote (9.5\%), \newline save (0.5\%) & 5 (day/user) & Poisson  & none & 0 (70\%), \newline 10-20 (29\%), \newline 50 (1\%) & none \\
\hline
City event \newline announcements & 2000-10000 & 2-5 out of: sale, concert, exhibition, outdoor, food, happy hour, market, sports, demonstration &  Ignore (80\%), \newline like (15\%), \newline save (4.5\%), \newline save\&go (0.5\%) & 0.1 (day/user) & Poisson  & 2-5 out of: sale, concert, exhibition, outdoor, food, happy hour, market, sports, demonstration & 0 (70\%), \newline 1-5 (29\%), \newline 10 (1\%) & Place: mostly city center, spread some around. \newline Time: mostly evenings and weekends. \\
\hline
Emergency \newline notification & 2000-10000 & none &  Read\&run (if close) (100\%) & 0.1 (day/user) & Poisson  & none & 100 (100\%) & Random \\
\hline
\end{tabular}
\end{scriptsize}
\end{center}
\label{tab:params}
\end{table*}%

\paragraph{Jodel-like Application}
The first example we explored is a Jodel-like application, where users post anonymous messages to share with everyone in their vicinity. Usually those messages are jokes, general questions about life and studies, thoughts, etc. The presented parameter set is based on our observation of messages over 3 days in the area of Bremen. This set of reactions is used throughout this paper. 

\paragraph{City Event Notifications}
This example refers to a service, where users post messages about interesting events happening in a city, such as concerts, sports events, markets, etc. The users enter the time and place of the event, as well as some keywords. The proposed set of keywords is rather small to keep the scenario manageable. The set of users' reactions include an ignore option, a like option, a save option an a special option to save and go to the event.

\paragraph{Emergency Notification}
This application targets the scenario, where a large-scale emergency happens, such as earthquake, large fire, tsunami, etc. The assumption is that everyone will read the message (for being informed or to be able to help) and, if in the danger area, will run away from it. The danger area itself is provided in the message.


\section{Mobility Model}\label{sec:mob}

Our mobility-enabled user behaviour model requires a mobility model, which is able to react to commands like "run away from X" and "schedule a move to Y at time Z". In principle, all existing mobility models can  be changed to accommodate such commands. In the following, we describe the implementation of R-SWIM (Reactive SWIM), which is a reactive extension of our Small Worlds in Motion (SWIM) implementation~\cite{udugama:2016swim}.


R-SWIM, similar to the SWIM, keeps a track of neighboring loations and visiting locations (i.e., remote loations), and decides on the next destination using the SWIM equation. This procedure is overridden when the User Behaviour Model decides on a specific reaction (as specified in Section~\ref{sec:user}). There are 2 types of reactions - \emph{Immediate} and \emph{Scheduled}. When an \emph{Immediate} reaction is expected, R-SWIM identifies a new location to travel to based on standard SWIM and initiates the movement. When a \emph{Scheduled} reaction is requested, R-SWIM queues the reaction to be executed when the appropriate time is reached. The appropriate time is decided by the mobility model based on the timing of the associated event and the distance to travel.

%
%
%
%
%


\section{Evaluation Metrics}\label{sec:metrics}

The real difference between traditional metrics and ours is what we understand under "delivered messages". Only messages, to which the user has reacted with more than "ignore" and were received on time are considered successfully delivered. All other messages, duplicates, etc. should be considered overhead. Otherwise we compute \textbf{delivery rate} and \textbf{delivery delay} as usual, taking into account the injection time of each message and its first delivery to the user. Furthermore, we compute \textbf{per user overhead} as a percentage between the number of messages this particular user has received over the number of her successfully received messages. All of the above mentioned metrics should not only be presented as means and confidence intervals, but also evaluated in terms of their \textbf{fairness distribution}. This is very useful to explore whether some users have been neglected in terms of data delivery and delay or overloaded with traffic.


\section{OMNeT++ Implementation}\label{sec:impl}

\begin{figure}[htbp]
\begin{center}
\includegraphics[width = \columnwidth]{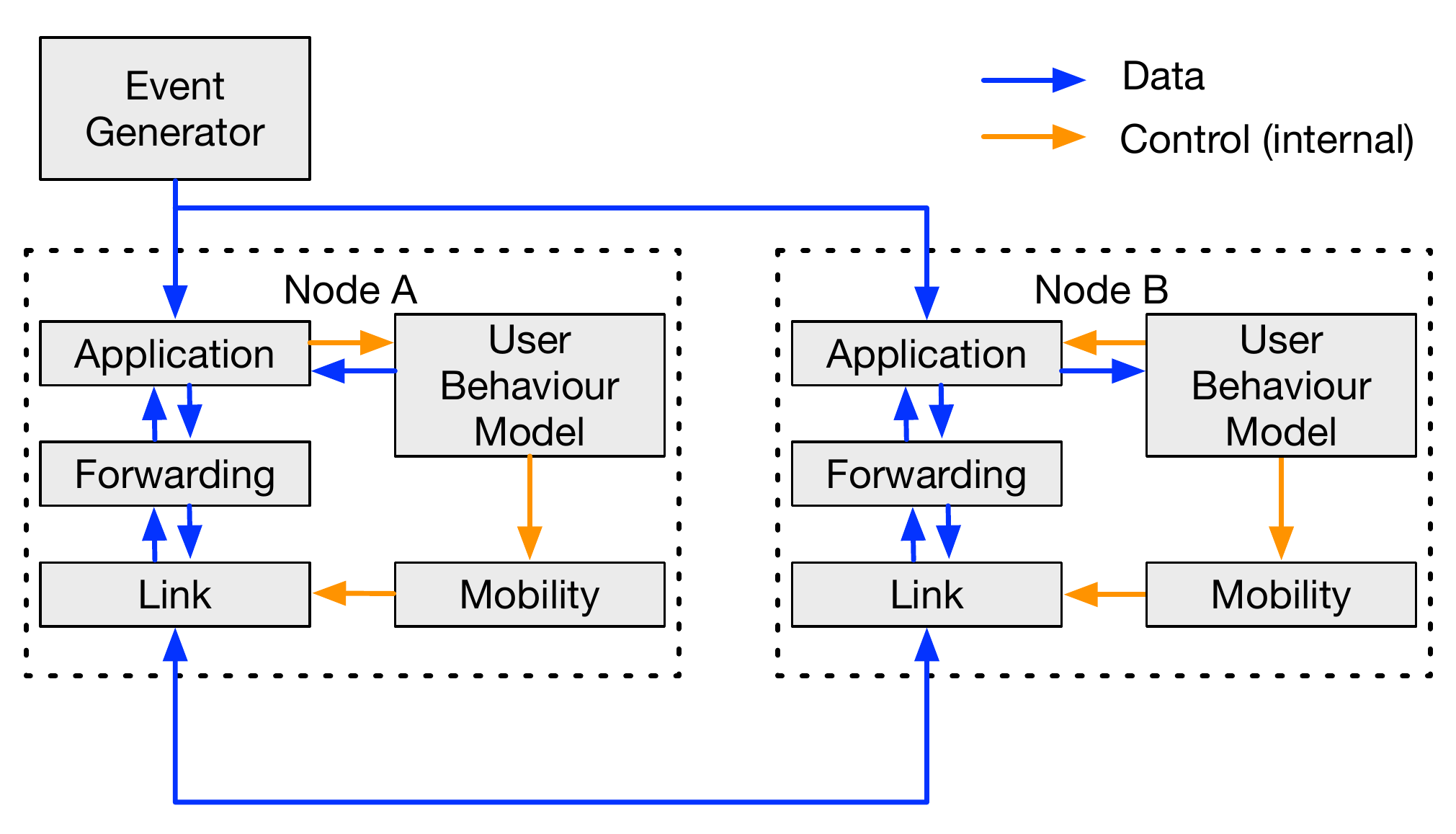}
\caption{Interactions between the individual modules in OPS.}
\label{fig:interact}
\end{center}
\end{figure}

We have implemented the above described models in OMNeT++. The structure of the implementation is depicted in Figure~\ref{fig:interact}. It is part of our Opportunistic Protocol Simulator (OPS) for OMNeT++. 

The behaviour of the models \textsc{UserBehavior, Mobility} follow the here presented models. \textsc{EventGenerator} is the module, which injects the messages into nodes. \textsc{Application} is rather a place-holder for more complex application scenarios and in this case simply forwards new messages to the \textsc{UserBehavior}. The other models are part of the OPS Framework.
%



\section{Conclusion and Future Works}\label{sec:conc}

In this paper we have mostly used our experiences with real users, social traces and expectations. The presented models show very promising results and are much more valuable for researchers than existing simulation models. However, we have not confirmed the user simulation parameters (e.g. the probability function to rate events) with real user traces. We are planning to do this in the immediate future.

Furthermore, we plan to extend the user behavior models with other applications and environments, e.g. weather forecast, transportation, etc. We are happy to cooperate with other research groups on these topics.

\bibliographystyle{IEEEtran}


\end{document}